\documentclass[sigconf]{acmart}

\usepackage{booktabs} 
\usepackage[ruled]{algorithm2e}

\acmConference[OpenSuCo'17]{OpenSuCo 2017}{November 2017}{Denver 
  , Colorado USA} 
\acmYear{2017} 
\copyrightyear{2017}

\newcommand{\octotiger}{OctoTiger\xspace}
\newcommand{\futurization}{Futurization\xspace}

\newcommand{\futures}{\texttt{future}s\xspace}

\newcommand{\channels}{\texttt{channel}s\xspace}

\usepackage{listings}

\usepackage{subfigure}

\begin{document}
\title[HPX -- An open source lib for Parallelism and Concurrency]{HPX -- An open source C++ Standard Library for Parallelism and Concurrency }

\author{Thomas Heller}
\affiliation{%
  \institution{Friedrich-Alexander University Erlangen-Nuremberg}
  \city{Erlangen}
  \state{Bavaria}
  \country{Germany}}

\author{Patrick Diehl}
\orcid{0000-0003-3922-8419}
\affiliation{%
 \institution{Polytechnique Montreal}
 \city{Montreal}
 \state{Quebec}
 \country{Canada}}

\author{Zachary Byerly}
\affiliation{%
  \institution{Louisiana State University }
  \city{Baton Rouge}
  \state{Louisiana}
  \country{USA}}

\author{John Biddiscombe}
\affiliation{%
    \institution{Swiss Supercomputing Centre (CSCS)}
  \city{Lugano}
  \country{Switzerland}}

\author{Hartmut Kaiser}
\affiliation{%
  \institution{Louisiana State University }
  \city{Baton Rouge}
  \state{Louisiana}
  \country{USA}}

\renewcommand{\shortauthors}{T.~Heller et al.}

\begin{abstract}
  To achieve scalability with today's heterogeneous HPC resources, we need a dramatic shift in our thinking; MPI+X is not enough. Asynchronous Many Task (AMT) runtime systems break down the global barriers imposed by the Bulk Synchronous Programming model. HPX is an open-source, C++ Standards compliant AMT runtime system that is developed by a diverse international community of collaborators called The Ste$||$ar Group. HPX provides features which allow application developers to naturally use key design patterns, such as overlapping communication and computation, decentralizing of control flow, oversubscribing execution resources and sending work to data instead of data to work. The Ste$||$ar Group is comprised of physicists, engineers, and computer scientists; men and women from many different institutions and affiliations, and over a dozen different countries. We are committed to advancing the development of scalable parallel applications by providing a platform for collaborating and exchanging ideas. In this paper, we give a detailed description of the features HPX provides and how they help achieve scalability and programmability, a list of applications of HPX including two large NSF funded collaborations (STORM, for storm surge forecasting; and STAR (\octotiger) an astrophysics project which runs at $96.8$\% parallel efficiency on $643$,$280$ cores), and we end with a description of how HPX and the Ste$||$ar Group fit into the open source community.
\end{abstract}
%
%



\maketitle

\section{Introduction}
The free lunch is over~\cite{free_lunch} - the end of Moore's Law means we have
to use more cores instead of faster cores. The massive increase in on-node
parallelism is also motivated by the need to keep power consumption in
balance~\cite{4476447}. We have been using large numbers of cores in promising
architectures for many years, like GPUs, FPGAs, and other many core systems;
now we have Intel's Knights Landing with up to 72 cores. Efficiently using that
amount of parallelism, especially with heterogeneous resources, requires a
paradigm shift; we must develop new effective and efficient parallel programming
techniques to allow the usage of all parts in the system.  All in all, it can be
expected that concurrency in future systems will be increased by an order of magnitude.

HPX is an open-source, C++ Standards compliant, Asynchronous Many Task (AMT)
runtime system. Because HPX is a truly collaborative, international project with
many contributors, we could not say that it was developed at one (or two, or three...)
universities. The Ste$||$ar Group was created ``to promote the development of
scalable parallel applications by providing a community for ideas, a framework
for collaboration, and  a platform for  communicating these concepts to the
broader community.'' We congregate on our IRC channel (\#ste$||$ar on freenode.net),
have a website and blog (stellar-group.org), as well as many active projects
in close relation to HPX. For the last 3 years we have been an organization in
Google Summer of Code, mentoring 17 students. The Ste$||$ar Group is diverse and
inclusive; we have members from at least a dozen different countries.
Trying to create a library this size that is truly open-source and C++ Standards
compliant is not easy, but it is something that we are committed to. We also
believe it is critical to have an open exchange of ideas and to reach out to
domain scientists (physicists, engineers, biologists, etc.) who would benefit
from using HPX. This is why we have invested so much time and energy into our
two scientific computing projects: STORM, a collaborative effort which aims to
improve storm surge forecasting; and STAR (\octotiger) an astrophysics project
which has generated exciting results running at $96.8$\% parallel efficiency on
$643$,$280$ cores. In addition leading the runtime support effort in the
european FETHPC project AllScale to research further possibilities to leverage
large scale AMT systems.

In the current landscape of scientific computing, the conventional thinking is
that the currently prevalent programming model of MPI+X is suitable enough to
tackle those challenges with minor adaptations~\cite{10.1109/MC.2016.232}.
That is, MPI is used as the tool for inter-node communication, and X is a
placeholder to represent intra-node
parallelization, such as OpenMP and/or CUDA. Other inter-node communication
paradigms, such as partitioned global address space (PGAS), emerged as well, which focus on one sided messages together
with explicit, mostly global synchronization. While the promise of keeping the
current programming patterns and known tools sounds appealing, the disjoint
approach results in the requirement to maintain various different
interfaces and the question of interoperability is, as of yet, unclear~\cite{dinan2013enabling}.

Moving from Bulk Synchronous Programming (BSP) towards fine grained, constraint based synchronization in order to limit the
effects of global barriers, can only be achieved
by changing paradigms for parallel programming.
This paradigm shift is enabled by the emerging
Asynchronous Many Task (AMT) runtime systems, that carry properties to alleviate the aforementioned
limitations. It is therefore not a coincidence that the C++ Programming Language,
starting with the standard updated in 2011, gained support for concurrency by
defining a memory model in a multi-threaded world as well as laying the
foundation towards enabling task based parallelism by adapting the
\lstinline!future!~\cite{future1,future2} concept. Later on, in 2017, based on those
foundational layers, support for parallel algorithms was added, which
coincidentally, covers the need for data parallel algorithms.
The HPX parallel runtime system unifies an AMT tailored for HPC usage
combined with strict adherence to the C++ standard. It therefore represents a
combination of well-known ideas
  (such as data flow~\cite{dennisdataflow,dennism98}, futures~\cite{future1,future2}, and Continuation Passing Style (CPS)) with new and unique
  overarching concepts.
The combination of these ideas and their strict application forms underlying
  design principles that makes this model unique. HPX provides:
\begin{itemize}
    \item A C++ Standards-conforming API enabling wait-free asynchronous execution.
    \item \futures, \channels, and other asynchronous primitives.
    \item An Active Global Address Space (AGAS) that supports load balancing through object migration.
    \item An active messaging network layer that ships functions to data.
    \item A work-stealing lightweight task scheduler that enables finer-grained parallelization and synchronization.
    \item The versatile and powerful in-situ performance observation, measurement, analysis, and adaptation framework APEX.
    \item Integration of GPUs with HPX.Compute~\cite{Copik:2017:USI:3078155.3078187} and HPXCL for providing a single source solution to heterogeneity
\end{itemize}
HPX's features and programming model allow application developers to naturally
use key design patterns, such as overlapping communication and computation,
decentralizing of control flow, oversubscribing execution resources and
sending work to data instead of data to work.
Using \futurization, developers can express complex data flow execution trees that generate
millions of HPX tasks that by definition execute in the proper order.
This paper is structured as follows:\\
In Section~\ref{sec::hpx} we summarize briefly the important features of HPX for parallelism and concurrency. The application of HPX in different high performance computing related topics are presented in Section~\ref{sec::applications}. In Section~\ref{sec::opensource} we review the aspect of open source software and community building. Finally, we conclude with Section~\ref{sec::conclusion}.
\section{HPX runtime components}
\label{sec::hpx}
\begin{figure}[htb]
    \includegraphics[width=0.45\linewidth]{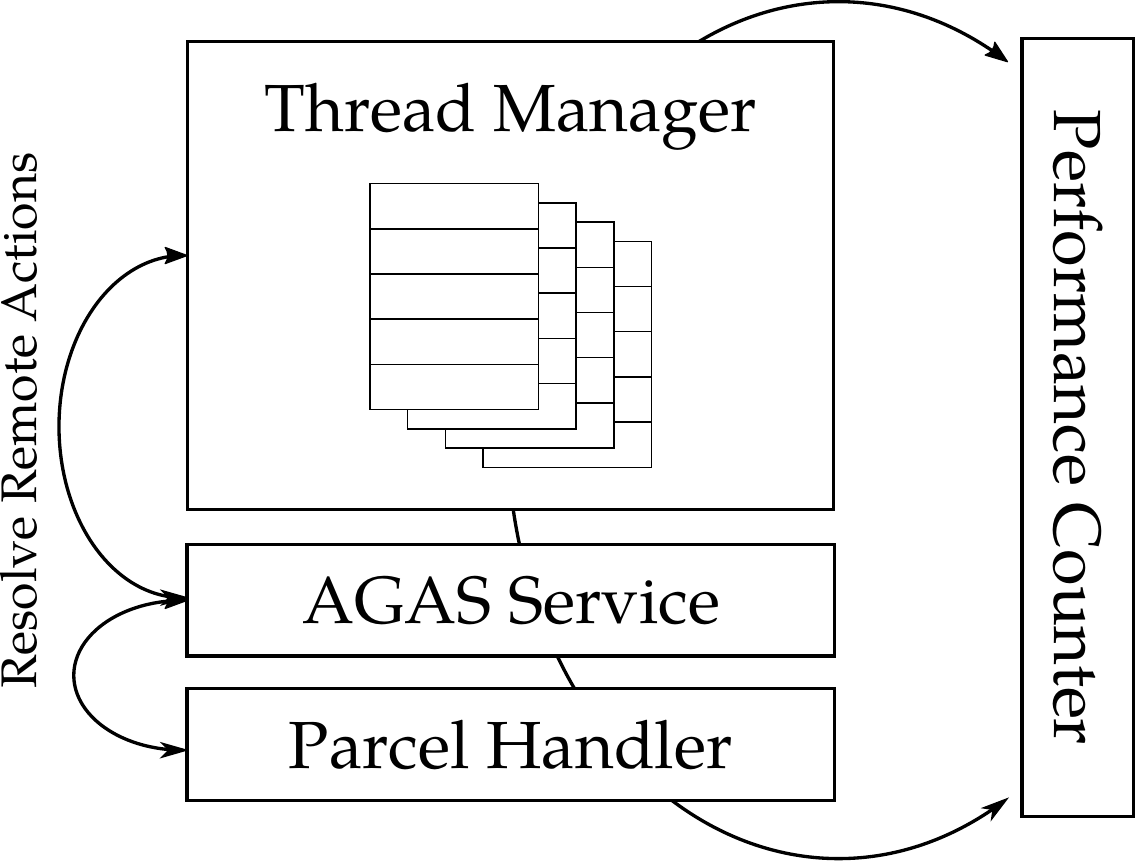}
    \vspace*{-1em}
    \caption[HPX Runtime System Components]{Overview of the HPX Runtime System components.}
    \label{fig:hpx:runtime_components}
\end{figure}

In this section we briefly give an overview of the HPX runtime system components with a focus on parallelism and concurrency. Figure~\ref{fig:hpx:runtime_components} shows five runtime components of the runtime system which are briefly reviewed in the following. The core element is the thread mamager which provides the local thread management, see Section~\ref{sec::hpx::threadmanager}. The Active Global Address Space (AGAS) provides an abstraction over globally addressable C++ objects which support RPC (Section~\ref{sec::hpx::agas}). In Section~\ref{sec::hpx::parcel} the parcel handler which implements the one-side active massaging layer is reviewed. For runtime adaptivity HPX exposes several local and global performance counters, see Section~\ref{sec::hpx::performcounters}. For more details we refer to~\cite{Kaiser:2015:HPL:2832241.2832244}.

\subsection{Thread manager}
\label{sec::hpx::threadmanager}
Within HPX a light-weight user level threading mechanism is used to offer the
necessary capabilities to implement all higher level APIs. Predefiend
scheduling policies are provided to determine the optimal task scheduling for a given
application and/or algorithm. This enables the programmer to focus on algorithms
instead of thinking in terms of CPU resources.
The predefined scheduling policies may be interchanged at runtime, additionally, 
new scheduling policies may be provided by the user for more application specific 
control. 
The currently available scheduling policies are summarized by:
\textbf{static: }the static scheduling policy maintains one queue per core 
while disabling task stealing alltogether; 
\textbf{thread local: }the thread local policy is currently the default policy.
This policy maintains one queue per core, whenever a core runs out of work, 
tasks will be stolen from neighboring queues, high priority queues also exist
to prioritize important tasks over others; 
\textbf{hierarchical: }the hierarchical policy maintains a tree of queues 
with the purpose of creating an hierarchical structure, 
while new tasks are always enqueued at the root, when one core fetches new work, the available tasks trickle down the hierarchy.

\subsection{Active Global Address Space}
\label{sec::hpx::agas}
After having the lightweight user level tasks available the next important module within HPX is the Active Global Address Space (AGAS). It's main purpose is to serve as the foundation to support distributed
computing and is related to other global address space attempts which are
represented by the PGAS family. Their main purpose is to hide explicit
message passing. By having each
object living in AGAS being represented by a GID (Global ID), the location of said
object does not need to be bound to a particular locality and leads to uniformity
in terms of independence whether a object is located remotely or local. In addition,
this independence allows for transparent migration between different process
local address spaces with AGAS being responsible for proper address resolution.
\subsection{Parcel}
\label{sec::hpx::parcel}
To support distributed computing, mechanisms for exchanging messages over a network
interconnect are necessary.
The current HPC landscape is dominated by two major programming paradigms.
The first is represented by MPI which offers
a wide variety of different communication primitives for passing messages between
processes. It is best known for its two-sided communication primitives where
the sending side issues a receive operation which contains the buffers, size of
the buffers, a tag and the destination. The destination has to proactively post
the receiving part of this operation. Those operations are extended by
asynchronous versions. In the HPX programming model, neither of those previously discussed forms of
communication seems to be appropriate. While having a versatile form of invoking
light weight threads locally, it seems natural that this should be extended in
such a way to allow for remote invocations of threads as well. This form of
communication is often referred to as RPC or Active Messaging. An
Active Message in HPX is called a Parcel. The parcel is the underlying
concept that allows us to implement RPC calls to remote localities 
in a C++ fashion. 
That means that we require a arbitrary, yet type safe, number of arguments, 
as well as the means to return values from remote procedure invocations. 
This operation is one-sided, as the RPC destination does not need to actively
poll on incoming messages, and follows the semantics of a regular C++ member function
call. For more details we refer to~\cite{scaling_impaired_apps, biddi2017}.

\subsection{Performance counters}
\label{sec::hpx::performcounters}
One important feature that helps improve performance portability is the means 
to be able to intrinsically profile certain aspects of a program. 
To assist in the decision making process, an HPX intrinsic
performance counter framework has been devised, which is able to encompass HPX
internal performance metrics as well as being extensible so that 
application specific metrics may be supplied, 
or platform dependant hardware performance counters (eg. using PAPI).
The performance counters are readable via AGAS, by having each
counter registered with a specific symbolic name, making them accessible throughout
the system (i.e. from any node). For more details we refer to~\cite{grubel2016using}.

\section{Applications of HPX in high performance computing}
\label{sec::applications}
The concepts of parallelism and concurrency were successfully applied in following applications.

\paragraph{LibGeoDecomp N-Body Code}
Scientific computing simulations have generally been developed by
domain scientists (physicists, materials scientists, engineers, etc.),
even those who either currently use or would greatly benefit from
using HPC resources. LibGeoDecomp, a Library for Geometric
Decomposition, reduces the effort required to develop scalable,
efficient codes. Libgeodecomp handles the spatial and temporal loops,
as well as the data storage, while the domain scientist supplies the
code required to update one unit. Heterogeneous machines make this
even more difficult. A 3D N-Body code written in LibGeoDecomp was used
to compare the performance of the MPI backend and the HPX backend.
HPX showed perfect scaling at the single node level (~98\% peak
performance) and ~89\% peak performance using the Xeon Phi Knights
Corner coprocessor~\cite{heller2013using}.  Further improvements to
the communication layer~\cite{Kaiser:2014:HTB:2676870.2676883} have
improved the performance of this application,outperforming the MPI
implementation by a factor of 1.4 and achieving a parallel efficiency
of ~87\% at 1024 compute nodes (16384 cores).


\paragraph{C++ Co-array Semantics}
HPX has been used to design a high performance API, written in C++,
that allows developers to use semantics similar to Co-array Fortran to
write distributed code. The goal was to enable the creation a more
easily understandable SPMD-style applications, while adhering to the
C++ standard and taking advantage of cutting edge features of the
language. Results from performance measurements show that scalability
and good performance can be achieved, and with much better
programmability and development efficiency, especially for domain
scientists~\cite{Tan:2016:ECC:2935323.2935332}.

\paragraph{GPGPU support}
There is currently no usable standards-conforming library solution for
writing C++ code that is portable across heterogeneous architectures
(GPGPUs, accelerators, SIMD vector units, general purpose cores). We
have set out to provide a solution to this problem with
HPX. \cite{heller2017} describes the design and implementation based
on the C++17 Standard and provides extensions that ensure locality of
work and data. The well-known STREAM benchmark was ported and compared
to running the corresponding code natively. We show that using our
single source, generic, and extensible abstraction results in no loss
of performance.
A SYCL backend for this API has also been developed, allowing for
single-source programming of OpenCL devices in C++. A comparison using
the STREAM benchmark showed only minimal performance penalties
compared to SYCL for large arrays.

\paragraph{Linear Algebra Building Blocks}
Codes that make use of different threading libraries (OpenMP, Kokkos, TBB, HPX, etc.)
for on node parallelism cannot be easily mixed since they all wish to \textit{own}
the compute resources and composability of code is reduced when diffferent
runtimes are mixed. We have therefore begun the process of creating
linear algebra building blocks that can make use of vendor optimized libraries
(such as MKL, cublas) on a node, but be integrated cleanly with HPX to make 
higher level BLAS type functions that are commonly used in HPC applications. 
Current benchmarks (Cholesky) show that the HPX versions perform as well as the
leading libraries in this area.

\paragraph{Storm Surge Forecasting}
Tropical cyclones pose a significant threat to both people and
property. The most damaging can cause tens of billions of dollars in
property damage, and the deadliest can kill hundreds of thousands of
people. Storm surge causes most of the
fatalities associated with tropical cyclones, but using computational
models it can be forecasted. ADCIRC \cite{luettich1992adcirc} is an
unstructured grid coastal circulation and storm surge prediction
model, widely used by the coastal engineering and oceanographic
communities. The unstructured triangular grid allows for seamless
modeling of a large range of scales, allowing higher resolution near
the coast and inland and lower resolution in the open ocean. Producing
reliable forecasting results enables emergency managers and other
officials to make decisions that will reduce danger to human lives and
property. Fast, efficient calculation of results is critical in this
role, and so the importance of high performance computing is
critical. We have used a novel approach ~\cite{Byerly2015} to update the Fortran MPI
implementation of a discontinuous Galerkin version of ADCIRC, DGSWEM,
to use HPX and LibGeoDecomp as a parallel driver, while still
retaining the original Fortran source code. LibGeoDecomp provides an
HPX data flow based driver which greatly mitigates problems of load
imbalance, exposes more parallel work, overlaps communication and
computation - all leading to decreased execution time. With HPX, we
are also able to run efficiently on Intel's Xeon Phi Knights Landing
on TACC's Stampede 2.

\paragraph{Astrophysics Research}
The astrophysics research group at Louisiana State University studies
the mergers of double white dwarf binary star systems using
hydrodynamic simulations. These mergers are believed to be the source
of Type Ia Supernovae, which due to the physics of the explosion,
always produce roughly the same amount of light. These "standard
candles" were used by two independent groups in the discovery of the
accelerating expansion of the universe, winning them the 2011 Nobel
Prize in Physics. Due to the relatively short timescale on which these
supernovae occur, the light from the binary star system leading up to
the merger has been rarely observed. So we rely on computer
simulations to give us insights into these mergers.  Previously, the
group at LSU used static uniform cylindrical mesh hydrodynamic code,
written in Fortran and parallelized with MPI to simulate these
mergers.  In a collaborative effort funded by the NSF, the group has
developed a highly efficient, Adaptive Mesh Refinement simulation
based on HPX, using the parallel C++ programming model which is now
being incorporated into the ISO C++ Standard. Futurization is used to
transform sequential code into wait-free asynchronous tasks. They
demonstrate a model run using 14 levels of refinement, which achieves
a parallel efficiency of 96.8\% on NERSC's Cori (643,280 Intel Knights
Landing cores) using billions of lightweight threads.

\section{HPX and Open source}
\label{sec::opensource}
Open source software (OSS) in high performance computing is already prevalent, i.\,e. all super computers in Top $500$ list released $2017$ run on Linux/Unix. More recently collaborative projects, like OpenHPC\footnote{\url{http://www.openhpc.community/}}, where vendors, educational institutions, and research organizations try to enable the complete HPC stack as OSS. One of their statements is they ``will provide flexibility for multiple configurations and scalability to meet a wide variety of user needs''. In the previous sections we argued that HPX can be used to address the scalability on multiple configurations. One example is the seamless integration of acceleration cards, like GPUs~\cite{Heller2016} and Xeon Phi, in the asynchronous execution graph. \\
Responding to requests from users, HPX core developers have a implemented several important features, including flexible but standards conforming parallel algorithms, IBVerbs and libfabrics parcelports, and thread priorities. As demonstrated by the ratio of closed tickets to total tickets generated by non-core developers, the team of HPX core developers is responsive to the requests of its users. HPX has a geographically decentralized team of core developers, each supported by seperate funding, and a much larger team of other developers contributing to not only to HPX but to other projects using HPX. Because of the exciting, cutting edge work being done, HPX attracts top C++ talent, many of which are students.

\begin{figure}[htbp]
\subfigure[Contributors per month. In the last to years we had around 10 different contributors per month.\label{fig::contributorspermonth}]{\includegraphics[width=0.45\textwidth]{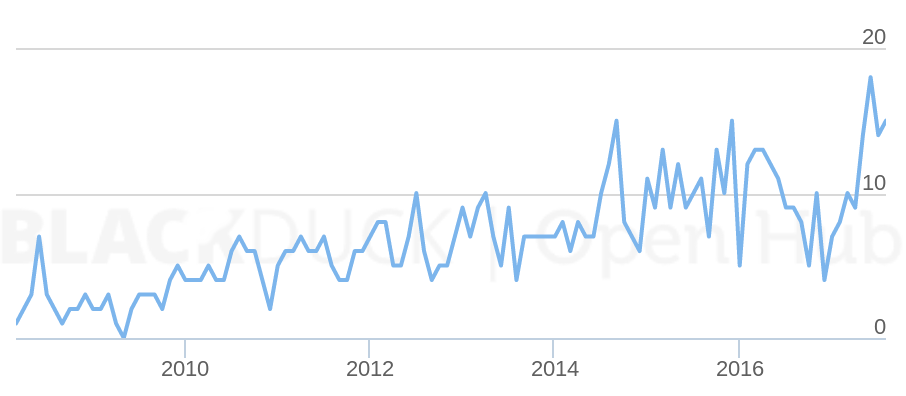}}\quad
\subfigure[Commit per month from 78 contributors. Note, that some of this contributors are GSoC students which contribute for three months during the summer. Therefore, in the summer the commits per month increase.
\label{fig::commitspermonth}]{\includegraphics[width=0.45\textwidth]{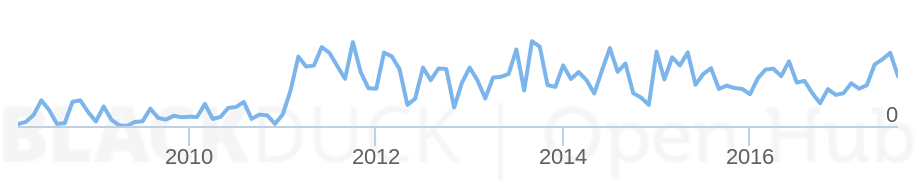}}
    \vspace*{-1em}
\caption{Commits and contributors from the beginning of 2008 when HPX was hosted on github. The picures were taken from openhub.}
\end{figure}

This section focus on the statistics with respect to open source software, see Table~\ref{tab::stats}. The complete source code of HPX is hosted on github\footnote{\url{https://github.com/STEllAR-GROUP/hpx}} ($702$ stars so far) and contains $598093$ lines of code where $465869$ are C++ code\footnote{\url{https://www.openhub.net/p/stellar-hpx/}}. HPX's stable version is $1.0$ containing $15$ releases. HPX is licensed under Boost Software License (Version $1$.$0$) which conforms 
to the Free/Libre Open Source Software (FOSS) concept. Continuous integration is done for Linux  with CircleCI and for Windows with AppVeyor.
\begin{table}[!htb]
\begin{tabular}{p{0.25\linewidth}|p{0.55\linewidth}}
Attribute  &  \\
\hline
License & Boost Software License, Version $1$.$0$ \\
\hline
Version  & $1$.$0$ with $15$ releases \\
\hline
Commits & $18431$ from $78$ contributors \\
\hline
Lines of Code & $465869$ (pure C++) and  $598093$ (all files)  \\
\hline
Continuous \mbox{Integration} &  CircleCI and AppVeyor \\
\hline
\end{tabular}
\caption{Statistics of HPX's source code and community}
\label{tab::stats}
    \vspace*{-1em}
\end{table}
Another aspect of open source software is its community; HPX is developed under the umbrella of Ste$\vert\vert$ar group which includes $78$ fellows from all over the world who contributed $17136$ commits. Figure~\ref{fig::contributorspermonth} shows that in the last two years approximately $10$ people contributed to HPX each month. Since $2014$ the Stellar group was accepted as an organization for Google Summer of Code (GSoC). In the last three summers $17$ students contributed valuable features to HPX or application using HPX. Figure~\ref{fig::commitspermonth} shows the commits per month since $2008$ when HPX was hosted on github. Since $2014$ the commits per month increase during the summer due to contributions of the GSoC students.

\section{Conclusion and Outlook}
\label{sec::conclusion}
On the latest and future super computers effective and efficient parallel
programming techniques are necessary to address the heterogeneity of the system.
In addition, the relevance of an open source solution for a HPC stack, like the
OpenHPC project, is developing. With HPX as an open source C++ standard library
for parallelism and concurrency both of these features are addressed. First, we
provide a parallel, AMT runtime system tailored to, but not limited to, HPC usage.
One focus is the strict adherence to the C++ standard, where well-known concepts, 
i.\,e. futures, are implemented, as well as contributing new
innovative ideas like Continuation Passing Style programming and dataflow to the
broader C++ Community. 
Second, HPX can be used to write applications for heterogeneous many core systems,
without changes to the code -- with HPX.Compute~\cite{Copik:2017:USI:3078155.3078187} 
we provide a unified model for
heterogeneous programming using CUDA and SYCL to provide a single source
solution to heterogeneity. With HPXCL we provide a means for application developers 
to write GPU kernels (CUDA and OpenCL) which can be compiled and run asynchronously 
on arbitrary devices in a heterogeneous system.

We have shown the successful application of HPX in geometric decomposition, storm
surge forecasting, linear algebra, astrophysics, crack and fracture mechanics,
and computational fluid dynamic. As one example, HPX showed perfect scaling for
the three dimensional $N$ body benchmark within LibGeoComp  at the single node
level ($98$\%) and ($89$\%) peak performance on the Xeon Phi Knights Corner
coprocessor and could outperform MPI implementation by a factor of $1.4$. For
the astrophysics study of merging double white dwarf binary star systems, a
parallel efficiency of $96.8$\% on NERSC's Cori using $643$,$280$ Intel Knights
Landing cores  was demonstrated using billions of lightweight threads.

\bibliographystyle{ACM-Reference-Format}
\bibliography{bibliography}

\end{document}